\documentclass[conference]{IEEEtran}
\IEEEoverridecommandlockouts
\usepackage{cite}
\usepackage{amsmath,amssymb,amsfonts}
\usepackage{algorithmic}
\usepackage{graphicx}
\usepackage{textcomp}
\usepackage{xcolor}
\def\BibTeX{{\rm B\kern-.05em{\sc i\kern-.025em b}\kern-.08em
    T\kern-.1667em\lower.7ex\hbox{E}\kern-.125emX}}
\begin{document}

\title{Real-time Coherency Identification using a Window-Size-Based Recursive Typicality Data Analysis
\thanks{This work is supported by the Brazilian agencies CNPq (grant
170100/2018-9), São Paulo Research Foundation (FAPESP) (grants
2016/08645-9, 2018/07375-3, 2018/20104-9 and 2019/10033-0), Engie (grant PD-00403-0053/2021).}
}

\author{\IEEEauthorblockN{
Lucas Lugnani}
\IEEEauthorblockA{\textit{University of Campinas}\\
Campinas, Brazil \\
lugnani@dsee.fee.unicamp.br}
\and
\IEEEauthorblockN{
Daniel Dotta}
\IEEEauthorblockA{\textit{University of Campinas}\\
Campinas, Brazil \\
dottad@unicamp.br}
\and
\IEEEauthorblockN{
Mario R. A. Paternina}
\IEEEauthorblockA{\textit{University of Mexico}\\
Mexico City, Mexico \\
mra.paternina@fi-b.unam.mx}
\and
\IEEEauthorblockN{
Joe Chow}
\IEEEauthorblockA{\textit{Rensselaer Polytechnique Institute}\\
Troy, USA \\
chowj@rpi.edu}
}

\maketitle

\begin{abstract}
This work presents a data-driven analysis of minimal length necessary for coherency detection considering a recursive form of the typicality-based Data analysis (TDA). It proposes a methodology that encloses the observation of the variance of the typicality ($\tau$) to asses the minimal window length necessary to determine the coherent buses, where the properties of the TDA approach and the groups of buses are iteratively calculated at every new data point sampled. Once the variance of each group reaches a certain value, the minimal window length is determined. Besides, this method preserves the TDA characteristics of using exclusively measurements, not requiring pre-determination of number of groups, group centers or cut-off constants. The method is applied to the well know 2-area Kundur test system, allowing to corroborate its effectiveness and draw conclusions regarding minimal window length dependence on the slowest inter-area mode.
\end{abstract}


\section{Introduction}
Coherency analysis is an important tool for power system operators, as complex models can be reduced to simpler equivalent ones and important parameters together with security margins can be verified. The dependence on models for coherence detection can be a critical factor, as power systems become more and more complex caused by the integration of inverter-based generation (IBG), making the model-based methods slower and more uncertain. Fortunately, power systems are also becoming better observed thanks to the installation of phasor measurement units (PMUs) that in turn, forming wide-area monitoring systems (WAMS) which provide a  high-dimensionality data set for purposes of identifying coherency and other features of the operators' concern. Among the opportunities provided by reduced models, response to ringdown disturbances are accordingly used for tuning controllers in areas or sub-areas belonging to the same or different bulk interconnected power grids.

Different actions taken by operators after large disturbances are dependent on the knowledge of coherent groups, such as coordinated control \cite{Messina98}. Besides the knowledge of coherent groups, it is also important that this information is provided as fast as possible, so the action taken by the operator is as efficient and minimal as possible. To this extent, it is crucial to determine the coherent groups with the help of little data as possible, that is, with the shortest window of measurements possible.

Investigations usually consider a fixed window length between 10 and 20 seconds, aiming to accommodate two periods of the slowest modes and additional time due to the transitory phenomena following a disturbance ~\cite{khalil2015dynamic,Lin2017CoherencyRenewables,paternina2018identification,Lin2018_CoherencySpectralClustering,lin2018wams,LiuRobustSeparation2020,Lugnani2021TDA}. These works briefly discuss the applicability of the methods for smaller windows, encouraging to the researcher to invest more efforts and discussion in providing solutions for such reductions.

In~\cite{SunUnifiedSeparation2011}, the authors propose a window length ranging from 20 to 60 seconds to accommodate all inter-area modes and also to perform controlled power system separation, taking into account three stages of such problems, namely: offline analysis, online analysis and real-time control stage. The first stage identifies elementary generation groups and devises control strategies for post-separation. The online stage performs modal analysis of synchrophasors for predictive separation points in the system, whereas the real-time control stage calculates a separation-risk index (SRI) based on time-synchronized measurements for finding the best time of separation. Meanwhile this method is comprehensive, other applications require smaller time windows since the higher penetration of IBG in the system has diminished the available time for operators' response.

The authors in~\cite{NaglicCoherencyWindowLength2020} propose a stability criterion-based method to  determinate the smaller window length, where the angular deviation among two recent measured vectors of generators' distances. This is calculated by their cosine dissimilarities and compared with heuristically-defined thresholds. If the value of the distance among vectors is smaller than the threshold, the window length is found. The authors claim that the window length is affected by the threshold choice that it has a range from 0.01 up to 0.1, where thresholds closer to the lower value present higher resolution of the groups, whereas values closer to the upper value exhibit more robustness to temporal changes.


The remainder of this work is organized in the following. Section II presents fundamental concepts of coherency and empirical data analysis required for the methodology presented in Section III. Section IV shows the application of the methodology to the Kundur two-area system, unveiling the characteristics and advantages of the method. Finally, Section V discusses the key findings and future works to enhance the method.

\section{Fundamentals}

\subsection{Coherency}

Coherency is defined as the behavior of generators swinging together in response to a disturbance~\cite{chow13Coherency}, with minimal distances between their responses. This swing behavior can be observed in angle ($\delta$) and frequency ($f$) measurements; likewise it can be derived from the linearized model analysis where the frequency of oscillatory modes and their participation factors indicate which generators are coherent to each other. Both approaches can be extended to non-generation buses and become less clear as the size of the system and the number of connections increase. In this scenario, a cut-off constant ($\gamma$) based on operator knowledge is used to determine the maximum distance that a bus must have to belong to a given group. This can be seen in ~\eqref{eq:1coherencydef} for frequency measurements:

\begin{equation}
    f_{i}(t)-f_{j}(t) \leq \gamma
    \label{eq:1coherencydef}
\end{equation}

For a window of angle measurements or frequency signals, there are several ways to approach the assessment of coherency. One can calculate the difference at every instant, making the tuning of the cut-off constant much more sensible to transitory periods and noise. Or a distance metric can  be also define as the best to represent the distance among two signals over a window length, making the tuning of $\gamma$ easier. However, the determination of the window length becomes crucial as the distance will also alter as the window expands or contracts. For the determination of coherency over ringdown disturbances, the incorrect choice of the window length may impact the determination of coherent groups. Usually, windows are determined considering the slowest known inter-area mode of oscillation of the system, where the length of the window is set to twice the period of that mode. However, this means windows of at least 2.2 seconds, as the fastest inter-area mode has a 1.1 second period, up to 10 seconds, considering the slowest modes. This delay may be prohibitive for applications that would require fast action, such as area coordinated damping controls~\cite{chow16}.

The distance $d(i,j,t)$ between the frequency measurements $f_{i}$ and $f_{j}$ can be measured in several ways, such as the absolute distance, Euclidean distance, the Frechet distance, the cosine dissimilarity, among many others and a combination of more than one~\cite{Lin2018_CoherencySpectralClustering}. This is also a dependent choice on the operator knowledge and can impact the clustering of buses, specially regarding the sensitivity of the consequential clustering method. It deals with with appropriate filtering of signal to increase the signal-to-noise ratio (SNR) and the proper adjustment of the size of the data set, that is, the window length.

\subsection{TDA}

The TDA approach is a data-driven method derived from empirical data analysis~\cite{angelov2017generalized} that approximates the probability mass function (PMF) of the data exclusively from the data themselves and a distance metric in accordance with the type of data measured, without any previous assumption of the distribution type (e.g., Gaussian, Logistic, Weibull, etc), nor number of modes, i.e., how many distribution means the data possess. To approximate the PMF, some properties must be calculated, but first a distance metric between the points must be defined. 

In~\cite{Lugnani2021TDA}, the TDA method is calculated for a fixed window $T$ of frequency measurements. Here, the recursive form will be presented, where the following properties are calculated at every new measurement $K$. As the interest is the coherency between measured signals, in other words the distance between signals, the Euclidean distance is used:

\begin{equation}
    d(i,j,t)=\sqrt{(f_{i}(t)-f_n)^2-(f_{j}(t)-f_n)^2}
    \label{eq:euclideandistance}
\end{equation}

\noindent where $f_{i}(t)$ and $f_{j}(t)$ are frequency measurements at the time instant $t$, $f_n$ is the system nominal frequency, and $d(i,j,t)$ is the distance metric between frequencies at the same time instant.

Next, three properties are calculated that lead to the value of the typicality $\tau$, which as it approximates the PMF. These properties are: \textit{i)} cumulative proximity; \textit{ii)} eccentricity and \textit{iii)} density. These properties are important as they are used for construction of the approximate PMF as equivalents to the statistical moments and at the clustering stage of the method, as the guarantee of the points, i.e. buses, belonging to a group is given by the Chebyshev inequality, where the eccentricity is used as a measure of anomaly within a group.

The first property, cumulative proximity $q(b_{i})$, is given as follows:

\begin{equation}
    q(b_{i})=\sum_{j=1}^{N}d^{2}(f_{i},f_{j})
    \label{eq:cumulativeproximity}
\end{equation}

\noindent where $N$ is the number of measuring points available. $q(f_{i})$ is a scalar that represents the \textit{total distance} of a point within a distribution based solely on the chosen metric or compound of metrics and $b_{i}$ is a point in the data set, which can be for example the norm of the frequency at bus $i$ for a given window of length $T$, or the vector of the norms of the distance between $f_{i}$ to every other bus. However, $q(b_{i}$ can also be recursively calculated at every instant $K$ as:

\begin{equation}
    q_{K}(b_{i}) = q_{K-1}(b_{i}) + d^{2}(b_{i},b_{j})
    \label{eq:generalrecursiveproximity}
\end{equation}

\noindent in which case we can consider $b_{i}$ solely as the measurement of angle or frequency at bus $i$ at instant $K$. For the Euclidean distance, \cite{angelov2019empirical} shows that the recursive proximity can be calculated as:

\begin{equation}
    q_{K}(b_{i}) = K(\|b_{i}-\mu_{K}\|^{2} + X_{K} - \|\mu_{K}\|^{2})
    \label{eq:euclideanrecursiveproximity}
\end{equation}

\noindent where $\mu_{K}$ and $X_{K}$ are the means of the set ${b}_{K}$ and ${b^{T}b}_{K}$, respectively, and both of them can be updated recursively as follows:

\begin{equation}
    \mu_{K} = \frac{K-1}{K}\mu_{K-1} + \frac{1}{K}b_{\forall i, K}
    \label{eq:meanofpoints}
\end{equation}

\begin{equation}
    X_{k} = \frac{K-1}{K}X_{K-1} + \frac{1}{K}\|b_{\forall i, K}\|^{2}
    \label{eq:meanofmatrix}
\end{equation}

Once this property is calculated recursively, all following properties can be updated at every new acquired measurement, in contrast to~\cite{Lugnani2021TDA}, where the properties are calculated once for the whole batch of measurements from the moment of disturbance up to $T= 10$. This recursive approach will become beneficial when we calculate the typicallity as $K$ increases and we observe its values reaching stability earlier than $T = 10$.

The second property is the eccentricity, which is a measurement of anomaly within the data set. Here, we show the normalized form of the eccentricity, $\epsilon$:

\begin{equation}
    \epsilon_{K}(b_{i}) = \frac{2q_{K}(b_{i})}{\frac{1}{K}\sum_{j=1}^{N}q_{K}(b_{j})}
    \label{eq:eccentricity}
\end{equation}

For the case where the distance matrix is Euclidean, the eccentricity can be calculated as:

\begin{equation}
    \epsilon_{K}(b_{i} = 1 + \frac{\|b_{i} - \mu_{K}\|^{2}}{\sigma_{K}^{2}}
    \label{eq:recursiveeccentricity}
\end{equation}

\noindent where $\sigma_{K}$ is the standard deviation, $\sigma_{K} = \sqrt{X_{K} - \mu_{K}^{2}}$

The normalized eccentricity is a very important metric because it indicates the points that are away from the peak of the data distribution. Hence, it can be used to find the tails of each mode in the distribution, or in our case, the buses that are borderline part of a given coherent group. If we recall the Chebyshev inequality for the Euclidean distance:

\begin{equation}
    P(\|\mu_{K} - b_{i}\|^{2} > n^{2}\sigma_{K}^{2}) < \frac{1}{n^{2}}
    \label{eq:generalChebyshev}
\end{equation}

\noindent where $n$ is the number of times the standard deviation away from the global mean  is being analyzed for $b_{i}$. Using the standardized eccentricity, the Chebyshev inequality can be reformulated of the form:

\begin{equation}
    P(\epsilon_{K}(b_{i}) > n^{2}+1) < \frac{1}{n^{2}}
    \label{eq:TDAchebyshev}
\end{equation}

With this expression, it can be said that, there is a smaller than $\frac{1}{9}$ probability of $\epsilon_{k}(b_{i} \geq 10$, for $n = 3$, which is a widely used condition for anomaly/tail detection. Furthermore, if the data distribution is Gaussian (which is not imposed by the TDA method), the probability of $\epsilon_{k}(b_{i} \geq 10$, for $n = 3$ is less than 0.3$\%$. This property is crucial for the clustering of buses without dependence on operator knowledge for setting a cut-off constant $\gamma$.

Next, we introduce the third property, the data density $D_{K}$, which is calculated as:

\begin{equation}
    D_{K}(b_{i}) = \frac{1}{\epsilon_{K}(b_{i})}
    \label{eq:density}
\end{equation}

Data density is the inverse of the eccentricity and data points that are closer to the mean have higher density values. The value of the data density evaluated at a particular data sample indicates how strongly this particular data sample is influenced by the other data samples in the data space due to their mutual proximity and attraction. It is also inversely proportional to the square distance between these two data samples.

The last calculated property is the typicality $\tau$, given as:

\begin{equation}
    \tau_{K}(b_{i}) = \frac{D_{K}(b_{i})}{\sum_{j=1}^{N}D_{K}(b_{j})}
    \label{eq:typicality}
\end{equation}

Analogously to the PMF, all points $b_{i}$ have $\tau(b_{i}$ within $(0,1]$, and the sum of all typicalities of the points of the distribution is equal to 1. However, since PMF is imposed to the data, it can have non-zero values for infeasible variable values (e.g., negative frequency), because characteristics of the variables are assumed prior to the data set. The points with higher typicality are the ones closer to the peak of the distribution, similar to PMFs like say, the tip of the bell curve for Gaussian distribution.

Once the properties are calculated, the clustering of buses is performed. With the recursive form of the TDA method, its properties and the clustering is done at every new measurement. The clustering process of the TDA method and more details of the method can be seen in~\cite{Lugnani2021TDA}. Next, the methodology for calculating the window length is presented, using the TDA method. 

\section{Methodology}

\begin{figure}[!tb]
    \centering
    \includegraphics[width=\columnwidth]{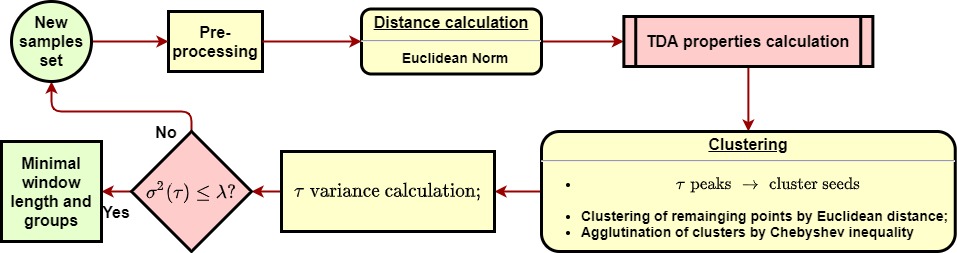}
    \caption{Recursive TDA methodology}
    \label{fig:methodolgoy}
\end{figure}

The proposed methodology is presented in Fig.~\ref{fig:methodolgoy}. It is calculated for every new batch of PMU measurements received until the point where the variance condition is attended. The minimal data sample is $5 \times N$ due to the filtering process, where $N$ is the number of PMUs available.

The first stage of the methodology is a pre-processing step to increase the SNR. As frequency measurements of PMUs are derived from voltage phasor angle measurements unrealistic spikes due to non-electromechanical phenomena may appear in both angle and frequency signals. To remove this effect, the first filter applied to the signals is a moving median filter, with a 5-sample window. Next, the DC offset is removed and the resulting signal is detrended with the dynamics separation algorithm~\cite{lackner2020voltage}.

The second step is the calculation of the distance, using~\eqref{eq:euclideandistance} to form a metric of the measurements distribution. With the points and their distances, it is possible to the calculate the TDA cumulative proximity ($q(b_{i})$~\eqref{eq:cumulativeproximity}), normalized eccentricity ($\epsilon_{K}(b_{i})$~\eqref{eq:eccentricity}), density ($D_{K}(b_{i})$~\eqref{eq:density}) and typicality ($\tau_{K}(b_{i})$~\eqref{eq:typicality}) properties. Starting at the second iteration of the methodology, $q(b_{i})$ and $\epsilon_{K}(b_{i})$ can be recursively calculated using~\eqref{eq:euclideanrecursiveproximity},~\eqref{eq:meanofpoints},~\eqref{eq:meanofmatrix} and~\eqref{eq:recursiveeccentricity}, respectively.

The third step is the clustering algorithm which uses the Chebyshev algorithm as a cut-off proxy in substitution of an user dependent constant. The clustering algorithm firstly ranks the points' typicalities by their Euclidean distances, starting from the highest typicality value. This creates a global distribution of typicalities based on their proximity and peaks of typicality are formed, if the distribution is multi-modal, indicating the existence of those modes. The peaks in the typicalities distribution are addressed as seeds of clusters $C_{m}$. Each cluster seed/peak receives its closest points by Euclidean distance, where the mean and standard deviation of each cluster $C_{m}$ are  computed. If a point is equally distant from two clusters, the point is addressed to the most likely cluster by the Chebyshev criterion, using the eccentricity of the point. After all clusters are formed and their first statistical moments are known, each $C_{m}$ is compared via Chebyshev inequality for a tail of $3\sigma$ with their mean values and the highest typicality of each cluster. If their means are closer than $2\sigma$, the cluster with highest typicality agglutinates the other, repeating the process until the number of cluster remains the same. More details regarding this algorithm can be seen in~\cite{Lugnani2021TDA}.

At each iteration $K$, the variance of the typicalities $var_{k}$ is  also calculated as:

\begin{equation}
    var_{K}(C_{m}) = \frac{\sum_{i=1}^{M}(\tau_{i}-\overline{\tau})^{2}}{M}
\end{equation}

\noindent where $var_{K}(C_{m})$ is the variance of cluster $C_{m}$ at the instant $K$, $M$ is the number of points in $C_{m}$ and $\overline{\tau}$ is the mean of the typicalities at $C_{m}$. Finally, if $var_{K}(C_{m})$, for every cluster, remains unaltered, say $\lambda = 0.5 s$, then the method converged to the coherent groups/clusters, with a window of length $K$. Otherwise, the method is repeated for the next batch of samples at $t=K+1$.

\begin{figure}
    \centering
    \includegraphics[width=\columnwidth]{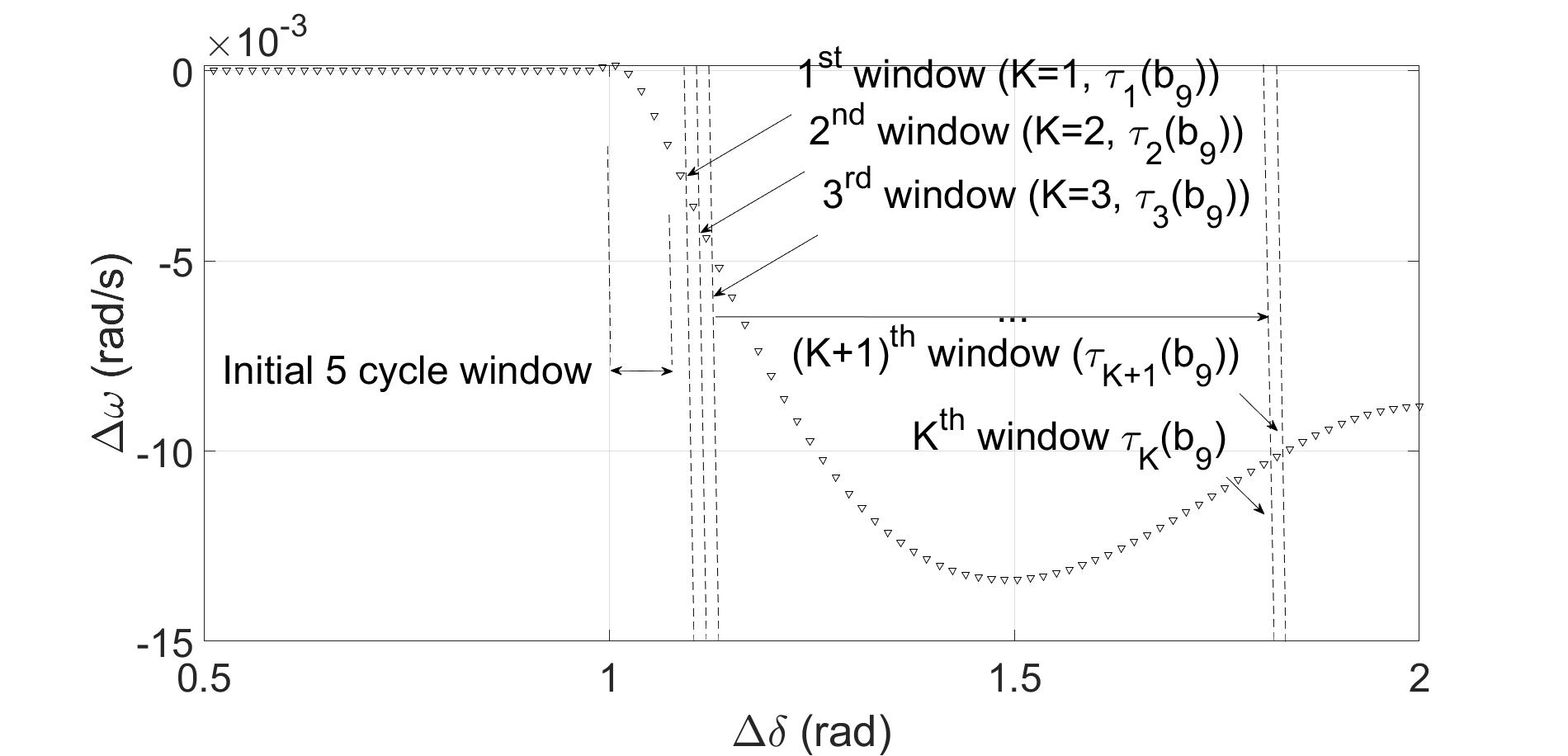}
    \caption{Size controlled window}
    \label{fig:ctrlwindow}
\end{figure}

The proposed methodology generates a size controlled window iteration process,  illustrated in Figure~\ref{fig:ctrlwindow} for bus 9 from Kundur 2-Area test system, where for new samples the method in Figure~\ref{fig:methodolgoy} is repeated until the variance criteria is satisfied.

Next, we show the application of the proposed methodology to the Kundur test system, and discuss the characteristics of the work.

\section{Results}

\begin{figure}[!tb]
    \centering
    \includegraphics[width=\columnwidth]{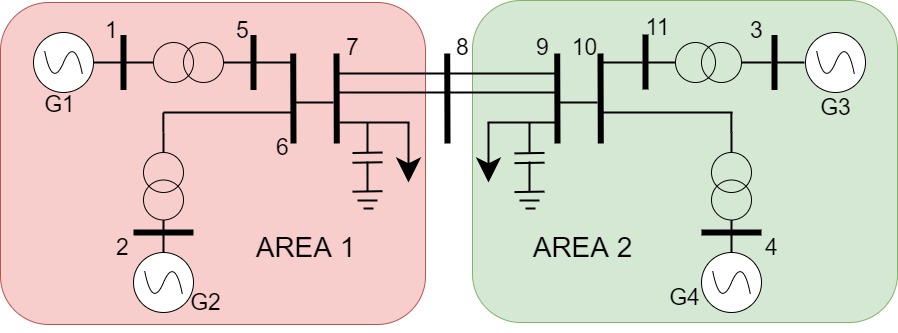}
    \caption{KundurTestSystem}
    \label{fig:kundurSys}
\end{figure}

The method is now applied to the 2-area Kundur system. The parameters of the system are the same as in~\cite{kundur94}, shown in Figure~\ref{fig:kundurSys}. This is a system with symmetrically well defined groups, due to its topology, and with a boundary bus (bus 8) that can be addressed to any group, depending on the tuning of the chosen coherency detection method. The groups are shown in Table~\ref{tab:groupsKundur}.

\begin{table}[]
    \centering
    \caption{Groups in the 2-area Kundur system.}
    \begin{tabular}{cc}
    \hline
    \hline
         & Buses \\
    \hline
        Group 1 & 1,2,5,6,7,8 \\
    \hline        
        Group 2 & 3,4,9,10,11 \\
    \hline
    \hline
    \end{tabular}
    \label{tab:groupsKundur}
\end{table}

\begin{figure}[!tb]
    \centering
    \includegraphics[width=\columnwidth]{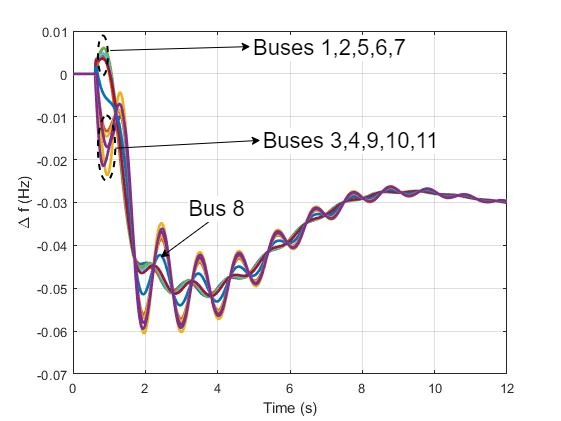}
    \caption{Frequency response of the Kundur test system to 100 MW step at bus 9.}
    \label{fig:frequncyresponse}
\end{figure}

The simulations were performed using the ANATEM software from CEPEL~\cite{ANAREDE17}, and the recursive TDA method was implemented with MATLAB R2018a, on an Intel Core i7-8850U 2.00 GHz processor with 8 GB of memory. To examine the proposed method, a 100 MW step is applied to bus 9 at 1 second which is the bus with highest load, in order to excite the ocillatory modes to be captured. The frequency response for all buses is presented in Fig.~\ref{fig:frequncyresponse}.

\begin{figure}[!tb]
    \centering
    \includegraphics[width=\columnwidth]{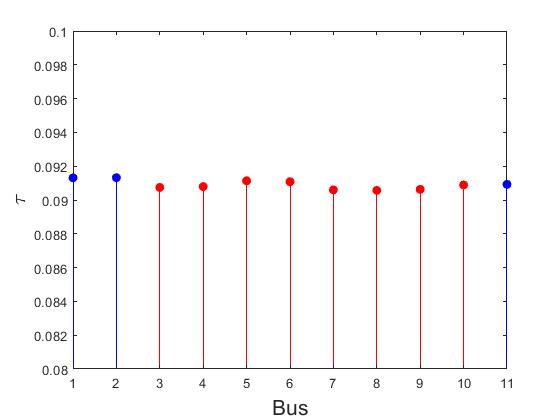}
    \caption{First set of typicalities.}
    \label{fig:FirstTausGroups}
\end{figure}

The methodology is applied to the frequency signals starting with 5 cycles due to the move median filter. The first set of calculated typicalities is shown in Fig.~\ref{fig:FirstTausGroups}, where the blue group is the group of generators 1 and 2, as reference. Note that the initialization of the method and the lack of information as electromechanical phenomena takes seconds to develop, the groups are incoherent, according to the coherency concept and the system topology.

\begin{figure}[!tb]
    \centering
    \includegraphics[width=\columnwidth]{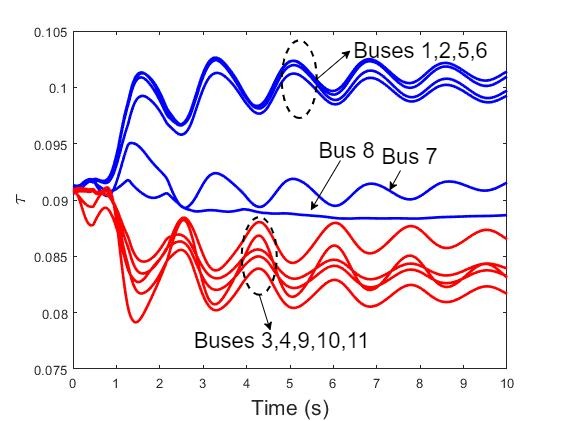}
    \caption{Evolution of typicalities.}
    \label{fig:TausOverTime}
\end{figure}

However, as time evolves and new samples are provided, the distribution of the data becomes more consistent with the coherent groups of the system, as shown in Fig.~\ref{fig:TausOverTime}. We can see that, even though the values of the typicalities oscillate, their values remain close to each other after a few seconds.

Figure~\ref{fig:TausOverTime} also shows the highest values of typicality in each group, namely buses 5 and 10. As the highest value of typicality represents the point closest to the mean of the distribution in such group, this bus can be interpreted as the center of the coherent group, since the mean of the distribution would represent the mean of the coherent response observed in frequency signals of the group. It is also interesting to note that the center buses of each group are not symmetrically equivalent, as in Area 2 the center bus is closer to the point of the fault.

Furthermore, Figure~\ref{fig:TausOverTime} reiterate the results corresponding to slow coherency clustering algorithm in~\cite{chowPowerBook20}, where eigenvectors associated with the inter-area frequency modes are computed and the mode shapes are used to form the slow coherency groups of generator and buses. However, due to the tail criterion of the TDA method, i.e. Chebyshev inequality, bus 8 is addressed to Area 1, whereas in~\cite{chowPowerBook20} the bus is left outside any group. Note that, depending on the disturbance, the areas using recursive TDA may change, as the window length, contrary to slow coherency method, which considers the power system linear model, hence the areas remain constant for the same operating condition.

\begin{figure}[!tb]
    \centering
    \includegraphics[width=\columnwidth]{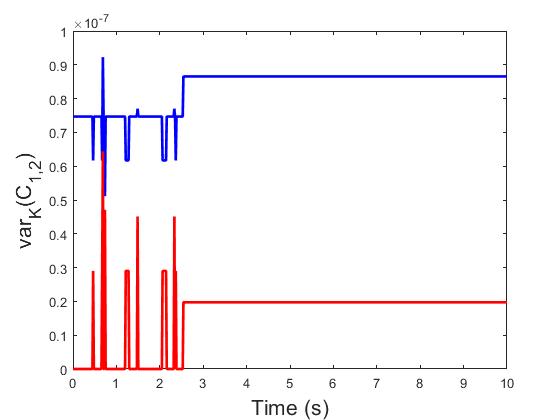}
    \caption{Variance of typicalities.}
    \label{fig:VarianceTau}
\end{figure}

This behavior can be clearly seen in Fig.~\ref{fig:VarianceTau}, where after about 2.5 seconds, the variance of the groups remains stable. Considering that the disturbance is applied at 1 second, the resulting difference is of 1.5 seconds. This is in accordance with the frequency of the inter-area mode of the system, which is of 0.545 Hz, with a period of approximately 1.8 second. This points to the fact that a window length of two times the period of the the slowest known inter-area mode of the system, as used in~\cite{Lugnani2021TDA,khalil2015dynamic} may be overzealous. For this methodology, the window length necessary would be of 1.5 seconds, plus the additional 0.5 seconds for assertion of the variance criterion, that is, a window length of 2 seconds.

It is important to note that the size of the window length may vary, according to the system or the system configuration itself. For instance, a bigger system with more modes, or a system with a slower mode may tend to have a different window length to accommodate the minimal information necessary in the signals.






\section{Conclusion}

This work has demonstrated that the recursive tipicality data analysis can be successfully implemented by using an adaptive window-size. Thus, the proposition of a recursive form of the TDA coherency detection method does not depend on an initial guess of the number of groups, its central points, neither an arbitrary cut-off constant. This is thanks to the recursive form removes the necessity of window length determination by the user, through the analysis of the variance of the typicalities within each group. The proposed method is applied to the Kundur test system to confirm its effectiveness and performance. Future works include the application of the method to real measurements and the implementation in simulated lab systems for online application. Additionally, the investigation of the Area center bus to represent the Area response will be pursued.









\bibliographystyle{./IEEEtran.bst}
\bibliography{./biblio.bib}




\end{document}